# Systems Development of a Two-Axis Stabilised Platform to Facilitate Astronomical Observations from a Moving Base*

James H. Hepworth
*Department of Mechanical Engineering*
*University of Cape Town*
Cape Town, South Africa
james.hepworth@alumni.uct.ac.za

Hendrik D. Mouton
*Department of Mechanical Engineering*
*University of Cape Town*
Cape Town, South Africa
hennie.mouton@uct.ac.za

*Abstract*—This project aimed to design, simulate, and implement a two-axis inertially stabilised platform (ISP) for use in astronomical applications. It aimed to approximate the stabilisation of a Meade ETX-90 3.5" compound telescope at low-cost using a mechanical assembly designed to geometrically and inertially model the telescope. A set of system specifications was developed to guide design decisions and to provide an analysis framework against which the performance of the implemented system was compared. The electro-mechanical structure of the ISP was designed and manufactured, the associated electrical systems were specified and configured, an image processing script capable of detecting and locating the centre of the Moon in a camera field-of-view was written, a complete simulation model for the system was developed and used to design various classical controllers for the ISP control system. These controllers were implemented on a STM32F051 microcontroller and a user interface was written in LabVIEW to facilitate intuitive user control of the system and perform datalogging of the system runtime data.

*Keywords— Inertial Stabilisation, Control Systems Design, Systems Modelling, Simulation*

## I. Introduction

Many examples of modern electronic and optical systems require inertial stabilisation for them to achieve their maximum performance[1], [2]. Inertially Stabilised Platforms (ISPs) are systems capable of isolating a sensor from its host by attenuating rotational disturbances coupled from the host to the sensor which result in a reduction in the sensor's performance.. ISPs aim to control the line-of-sight (LOS) between a sensor and a target. They perform two distinct operations; keeping track of the target as the sensor host and the target move in inertial space and attenuating rotational rate disturbances incurred to the sensor by host vehicle motion.

Observing a celestial object at high magnification through a telescope illustrates the problem clearly; even the smallest rotation of the telescope will cause, at the very least, blurring of the image observed, but is likely to cause a loss of the target altogether from the telescope field-of-view (FOV). A hobby-telescope affixed to the Earth is subject to this problem and at high magnification, celestial objects move quickly out of the FOV solely due to the Earth's rotation. The inclusion of an automatic target tracking and stabilisation control system to a telescope would provide a means to overcome this problem and to increase the robustness of the observation process: If the telescope was accidentally moved after observations had begun, the system would be able to keep the target in the centre of the FOV of the sensor. This would also allow for the telescope to be mounted on a moving base such as the deck of a ship or a moving motor vehicle.

Other papers have discussed the development of the dynamics models for gimballed mechanisms and factors associated with precise control of these systems including non-linear disturbances and structural effects, and presented control and compensation techniques to improve stabilisation performance [1]-[3]. This paper presents the design and implementation of all systems of a two-axis stabilised platform and automatic target tracker developed to facilitate the control of a Meade ETX-90 telescope to automatically track the Moon from a moving vehicle. Firstly, the main specifications of the system are presented before the design, implementation and tested performance of the ISP are summarised.

## II. Main System Specifications

The following specifications formed part of a broader set of technical design specifications and defined the overall design goals for the ISP system developed.

### A. Mechanical Specifications

    i.    The system should facilitate the inertial control of a 3.5" telescope.

    ii.   The system should allow for motion control about two control axes.

    iii.  The total system mass should be limited to a 9 kg.

### B. Operational Specifications

    iv.   The target tracker should have a tracking error of less than 0.25 mrad.

    v.    System jitter should be limited to 2 mrad under dynamic host conditions and 0.5 mrad under static host conditions.

*C. Constraints*

    vi.     Development costs should be less than R5000.00

## III. Systems Development

This section presents the development of the design of the ISP implemented in this project. The electro-mechanical assembly of the ISP system serves the purpose of mounting, constraining and controlling the motion of the telescope to be stabilised. Two-axis of control were deemed sufficient for the project as rotation of the telescope FOV about its centre was allowable during operation. This section begins with a description of the system dynamics used to model the gimballed mechanism employed by the ISP, before a description of the mechanical, electrical and software systems developed is given.

*A. System Model*

   *1) Axes and Angle Definitions*

Fig. 1 below shows the schematic of the gimballed mechanism used to model the ISP. In the figure, the gimbal facilitates rotation about the $z_b, z_g$ axis whilst the platform rotates about the $y_g, y_p$ axis. Here, the gimbals are modelled as rigid-bodies suspended on frictionless joints, however, later, in the system simulation shown in Section IV, inter-gimbal friction was modelled. The telescope is mounted on the platform. Three reference frames are defined using triads of orthogonal vectors, $\{x, y, z\}$ fixed to their hosts such that the frames rotate with the members to which they are affixed. These frames are defined as:

    i.     The Inertial reference frame, I, given by $\{x_i, y_i, z_i\}$,

    ii.    The Base fixed frame, B, given by $\{x_b, y_b, z_b\}$,

    iii.   The Gimbal fixed frame, G, given by $\{x_g, y_g, z_g\}$, and

    iv.   The Platform fixed frame, P, given by $\{x_p, y_p, z_p\}$.

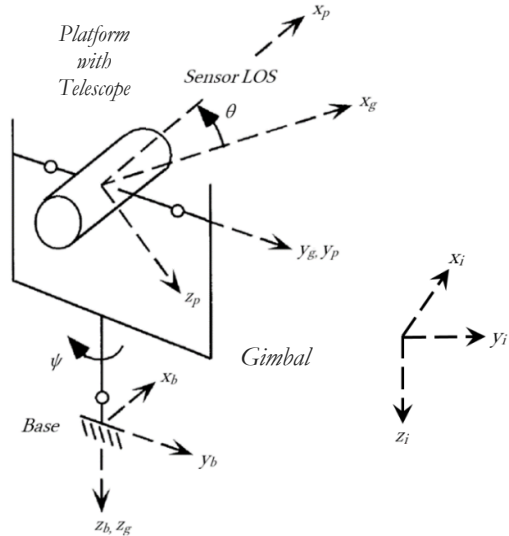

Fig. 1. Gimballed mechanism - Adapted from [4]

The angular positions of the frames relative to each other are given by the yaw angle (the angle between frames B and G), $\psi$, and the pitch angle (the angle between frames G and P), $\theta$. These are shown by Fig. 2 below.

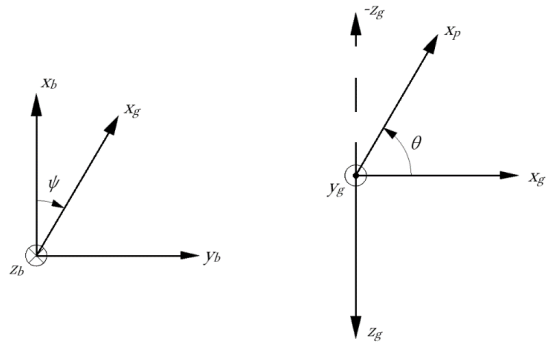

Fig. 2. Relative angle definitions

$\dot{\psi}$ is, therefore, the angular rate between the base and the gimbal, and $\dot{\theta}$ is, therefore, the angular rate between the gimbal and the platform. The definitions given above are used to represent a Euler sequence of the order yaw-pitch or $\psi$-$\theta$, from the base to the platform.

## 2) Kinematic Relationships

This section is heavily drawn from [5], but also draws from [4] and [6]. The two-axis gimbal system in discussion consists, ideally, of two rigid bodies free to rotate relative to each other. Gimbal-fixed and Platform-fixed co-ordinate systems, $\{x_g, y_g, z_g\}$ and $\{x_p, y_p, z_p\}$ (frames G and P) are attached to the gimbals and rotate with them. Frame B, $\{x_b, y_b, z_b\}$, is attached to, and rotates with the base on which the gimbals are mounted. The general equations of motion for the gimbals of the two-axis system are derived from the law of the conservation of angular momentum which, when defined relative to a moving coordinate system with angular momentum $\bar{H}$ and angular velocity $\bar{\omega}$, where the net moment $\bar{M}$ about the system is given by [7]:

$$\sum \bar{M} = \bar{\bar{I}} \cdot \dot{\bar{\omega}} + \bar{\omega} \times \bar{H} \tag{1}$$
$$\text{where } \bar{H} = \bar{\bar{I}} \cdot \bar{\omega}$$

$\bar{\bar{I}}$ is the tensor of inertia of the body to which the rotating frame is affixed. It follows then, that (1) can be used to develop the general equations of motion for the two-axis gimbal system as defined in Fig. 1 above. The platform will first be evaluated before evaluating the yaw-gimbal.

### a) Platform Equation of Motion

If the platform is suspended from its principal axes, the assembly has a tensor of inertia, $\bar{\bar{I}}_p$, and an angular rate, $\bar{\omega}_p$, such that:

$$\bar{\bar{I}}_p = \begin{bmatrix} I_{xxp} & 0 & 0 \\ 0 & I_{yyp} & 0 \\ 0 & 0 & I_{zzp} \end{bmatrix}$$

And,

$$\bar{\omega}_p = \begin{bmatrix} \omega_{xp} \\ \omega_{yp} \\ \omega_{zp} \end{bmatrix}$$

Applying (1) to the platform, the net external torque on the platform, $\bar{T}_p$, is given by:

$$\bar{T}_p = \bar{\bar{I}}_p \cdot \dot{\bar{\omega}}_p + \bar{\omega}_p \times \bar{H}_p$$

This gives the net external torque about the platform, $y_g, y_p$, axis, $T_{yp}$:

$$T_{yp} = I_{yyp} \dot{\omega}_{yp} + (I_{xxp} - I_{zzp}) \omega_{xp} \omega_{zp} \tag{2}$$

(2) defines the equation of motion for the platform for its rotation about the $y_g, y_p$ axis.

### b) Gimbal Equation of Motion

Applying (1), to the gimbal, which is assumed to be suspended from its principal axes, and has a tensor of inertia, $\bar{\bar{I}}_g$, and an angular rate, $\bar{\omega}_g$, the net torque on the gimbal, $\bar{T}_g$, is given by:

$$\bar{T}_g = \bar{\bar{I}}_g \cdot \dot{\bar{\omega}}_g + \bar{\omega}_g \times \bar{H}_g$$

$\bar{T}_g$ is the net torque about the gimbal, therefore, the external torque applied to the gimbal is given by $\bar{T}_{gp}$:

$$\bar{T}_{gp} = \begin{bmatrix} T_{xgp} \\ T_{ygp} \\ T_{zgp} \end{bmatrix} = \bar{T}_g + \bar{\bar{L}}_{PG}^T \cdot \bar{T}_p$$

Then,

$$T_{zgp} = (I_{zzg} + I_{zzp} \cos^2 \theta + I_{xxp} \sin^2 \theta) \dot{\omega}_{zg} \tag{3}$$
$$+ (I_{zzp} - I_{xxp}) \sin \theta \cos \theta \, \dot{\omega}_{xg}$$
$$+ \omega_{xg} \omega_{yg} (I_{yyg} - I_{xxg})$$
$$+ \omega_{xp} \omega_{yp} (I_{yyp} - I_{xxp}) \cos \theta$$
$$+ \omega_{yp} \omega_{zp} (I_{yyp} - I_{zzp}) \sin \theta$$
$$+ I_{zzp} \dot{\theta} \omega_{xp} \cos \theta$$
$$+ I_{xxp} \dot{\theta} \omega_{zp} \sin \theta$$

(3) defines the equation of motion for the platform for its rotation about the $z_b, z_g$ axis and together with (2) defines the motion of the double gimbal mechanism in terms of base motions and the external torques on each gimbal. They are used in the gimbal dynamics block shown in Fig. 10 of Section IV below.

*B. Mechanical Design*

The mechanical assembly designed in this project was ultimately required to mount the ETX-90 telescope and facilitate control of its inertial orientation. However, due to budget constraints early in the life of the project, it was decided to perform all initial systems development of the ISP without the purchase of the actual telescope, and to rather design a low-cost camera mounting platform to inertially and dimensionally model the ETX-90 and so facilitate the remainder of the systems being developed within the revised budget. To accomplish this an ETX-90 on loan was first modelled using SOLIDWORKS® and used to establish an estimate of its inertial and geometric properties. The estimated mass of the platform telescope mount was approximately 1.95 kg and had a tensor of inertia, $\bar{\bar{I}}_p$:

$$\bar{\bar{I}}_p = \begin{bmatrix} 0.0049 & 0 & 0.0008 \\ 0 & 0.0165 & 0 \\ 0.0008 & 0 & 0.0162 \end{bmatrix} \text{kgm}^2.$$

These parameters were then used to inform the design of the low-cost camera mounting system which was used in place of the telescope platform to facilitate systems development and testing over the remainder of the project. This telescope modeler closely represents the ETX-90 platform, geometrically and inertially, with an estimated mass of 1.90 kg and a tensor of inertia, $\bar{\bar{I}}_p$:

$$\bar{\bar{I}}_p = \begin{bmatrix} 0.0048 & 0 & 0.0009 \\ 0 & 0.0164 & 0 \\ 0.0009 & 0 & 0.0166 \end{bmatrix} \text{kgm}^2.$$

Following the design of the telescope modeller, the gimbal and mounting stand were designed to support the platform. Care was taken in the design of both the platform and the gimbal to ensure that the assemblies were suspended as close to their principal axes as possible so that the assumption made in the kinematics development regarding zero product of inertia terms was not invalidated. For both the EXT-90 and the telescope modeller, however, the product of inertia terms between the $x_p$ and $z_p$ axes were non-zero, however as these were only approximately 5% of the main moment of inertia terms, $I_{yyp}$ and $I_{zzp}$, the approximation of these terms to zero was deemed acceptable for the purposes of this project. The final design, shown in Fig. 3 ultimately had an estimated mass of 6.40 kg.

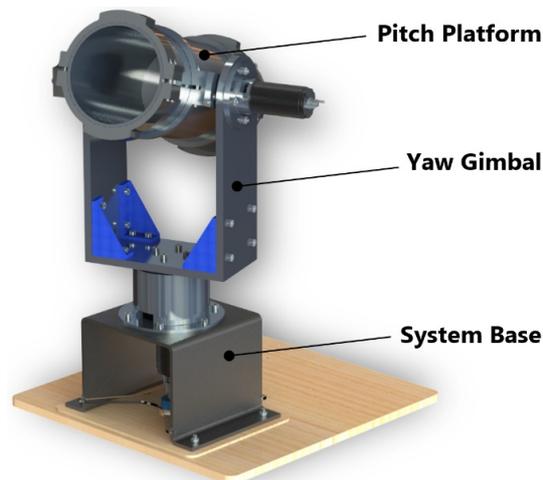

Fig. 3. Final design of the ISP and telescope modeller

*C. Electrical and Electronic Systems*

The electrical and electronic components specified and implemented in this project included the various feedback sensors required to measure the control parameters of the stabilised platform, the electrical systems associated with the actuation of the gimbal and the platform, and the controller hardware used to manage and control the overall system. In this application the inertial rotational rate of the platform LOS is the parameter whose desired value is to be controlled. Measurement of this rate was performed using an InvenSense MPU-9150 MEMS inertial measurement unit (IMU) which was able to measure the inertial rates of up to 250 °/s about three orthogonal axes and output its measurements over an I²C interface. The IMU had suitably low noise and non-linearity properties, high bandwidth and was available at a low cost. This sensor facilitated stabilisation control of the platform to be achieved. Secondary to the inertial rate sensor, for automatic target tracking to be implemented, a camera sensor was required to be specified. The image from this sensor would provide visual feedback of the target and enable automatic tracking through the stabilisation controller working in conjunction with an image processing algorithm used to locate the target in the camera's FOV. For this project a Raspberry Pi (RPi) Model 3 B computer and Raspberry Pi Camera v1.3 with a modified higher magnification lens were used to achieve the imaging and target location requirements of the system. The relative angles between the base and gimbal and the gimbal and the platform were measured using conductive plastic continuous rotation potentiometers
which were low-cost and had low rotational friction properties. Fig. 4 shows the implementation of these three sensor types on the ISP.

*1) Actuators*

Faulhaber 3257024CR 24V DC motors were chosen for use as the gimbal and platform actuators in this project and were driven by a dual channel, PWM controlled, MOSFET H-bridge.

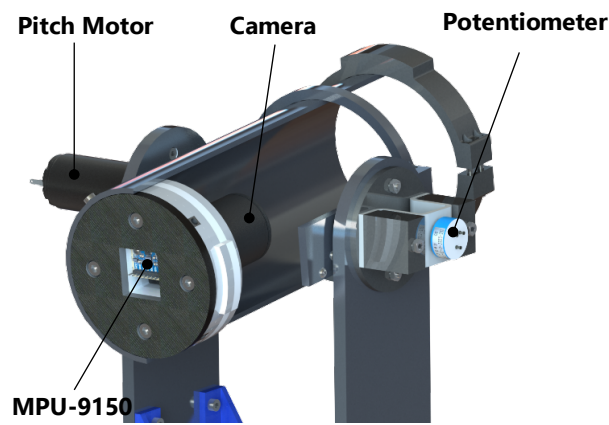

Fig. 4.  Mounting positions of the system sensors

These motors were directly coupled to the mechanics without the use of a gearbox to avoid the backlash induced disturbances and the inherent coupling of the base motion to the gimbals motion when gear drives are used [8].

*2) Control Hardware and Architecture*

The control hardware chosen for this project included an STM32F0 microcontroller (MCU) which was tasked with interacting with the various sensors and actuators to execute the control laws and communications tasks. In addition, a laptop computer was used to provide a user interface (UI) to facilitate supervisory control of the ISP and perform system datalogging tasks. These controllers were configured, along with the RPi, in a control architecture shown by Fig. 5.

*D. Software and Firmware Systems*

Three main software/firmware systems were developed for the ISP and used to implement the control architecture shown above: Embedded firmware, written in C, for the STM32 MCU which implemented the control laws and managed system communications tasks, a LabVIEW User Interface (UI) facilitating supervisory control operations, and an image processing script written in Python 2.7

The MCU was tasked with reading the angle pickoffs, communicating with the MPU-9150 IMU, managing comms with the RPi which sends target position data to the MCU, and running the control loops, which produce the actuation signals for the yaw and pitch motors. Additionally, the MCU is tasked with sending the system runtime data to the UI for data logging.

The RPi was tasked with processing the image produced by the RPi Camera to determine the pixel position of the target centroid, which is then sent to the MCU when the MCU initiates the data request. For this project, a simple colour recognition script was written which returned pixel positions of the centroid of the largest instance of the configured colour. This enabled laboratory testing of the system to occur easily with simulated targets and for simple reconfiguration to recognise the white Moon on the dark background of the night sky to be possible.

IV. SYSTEM SIMULATION AND CONTROLLER DESIGN

Following the systems development of the components of this project's ISP, this section describes the design of the various controller algorithms implemented on the control hardware. Namely, these include manual position controllers for the yaw and pitch channels, stabilisation controllers, and finally tracking controllers for the two channels. Together, these control loops facilitate the control of the ISP such that its orientation and motion may be controlled to a desired state under various operating conditions.

All controllers developed for the ISP were chosen to be classical controllers of type P or PI with compensation. Although modern and non-linear control methods have been shown to achieve superior stabilisation performance, classical controllers were chosen due to the prevalence of their use in the development of stabilised platforms [3], [8], [9]. Additionally, as this project represented the initial systems development of this ISP, classical controllers were the logical starting point for the development of its control system. All controller parameters designed in this section were determined using the Bode technique and system responses were evaluated under linear and specific saturated operating conditions representative of the expected operating environment of the ISP. All controllers discussed here, were designed in the continuous s-plane before being transformed to the z-plane using the bilinear transformation and implemented on the MCU using difference equations.

*A. Manual Position Controllers*

The manual position controllers serve the purpose of facilitating control of the initial orientation of the telescope before automatic target tracking and stabilisation are initiated. These, however, are not critical to the performance of the ISP and are not detailed further here.

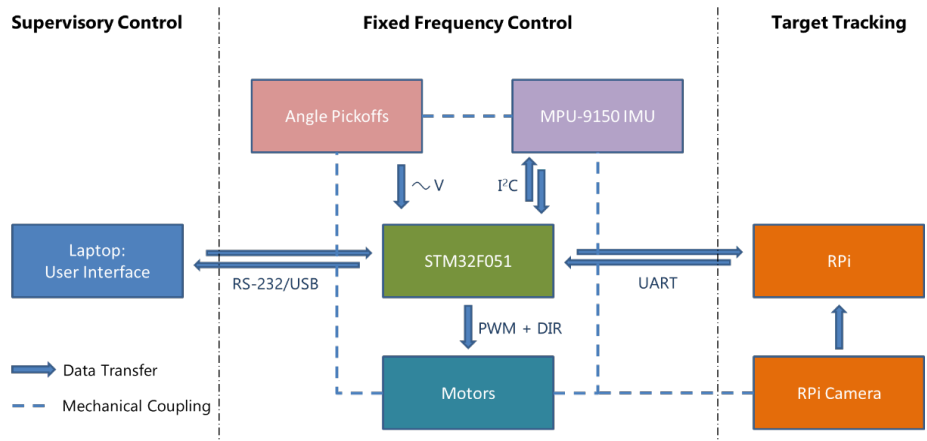

Fig. 5. Control hardware configuration

*B. Stabilisation and Tracking Controllers*

Typically, an ISP control system consists of two loops per axis configured in a cascade control structure. The outer loop is a low-frequency tracking loop used to keep the sensor pointing toward the target. It also has the objective of removing low-frequency parallactic motion between the camera or telescope and the target, and any possible drift in the rate loop. The inner loop is a high-frequency stabilisation loop used to control LOS rate and account for high-frequency disturbances [3], [8]. Fig. 10 shows the overall system model block diagram used for the design of these controllers.

In Fig. 6, the gimbal dynamics block implements the kinematic relationships between base motions and consequent external disturbance torques applied to the gimbals.

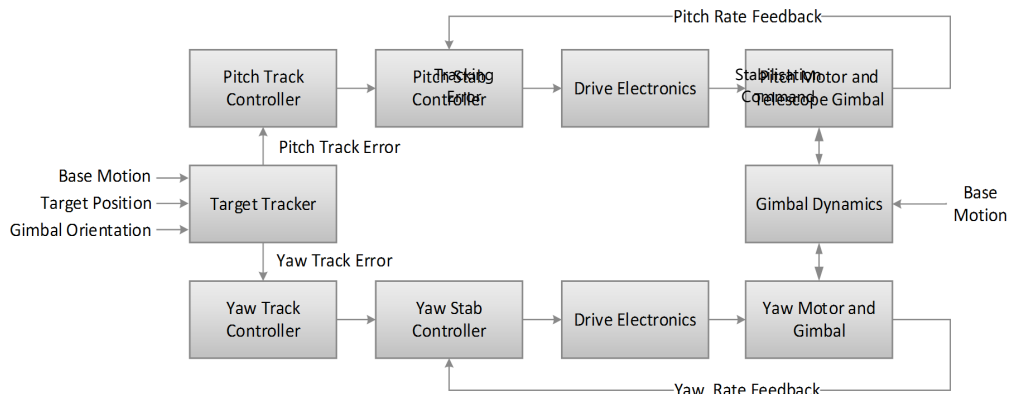

Fig. 6. System model block diagram

The target tracker block includes the relevant mathematical model required to calculate the yaw and pitch tracking errors from the inertial position of the target and the current inertial orientation and rates of the ISP. The yaw and pitch tracking controllers were modelled as shown by Fig. 7.

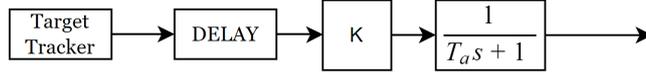

Fig. 7.   Tracking controllers

P controllers with a compensator with a single pole causing a phase lag at frequencies above 10 Hz were implemented as the tracking controllers. The delay block above models the delay caused by the image processing algorithm in identifying the target in the FOV. These controllers showed no resonance and had a low bandwidth of approximately 1 Hz which, however, was sufficient for the application as celestial bodies move slowly across the sky at frequencies far lower than 1 Hz. The yaw and pitch stabilisation controllers were modelled as shown in Fig. 8.

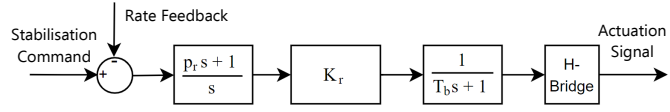

Fig. 8.   Stabilisation controllers

PI controllers with a compensator with a single pole causing a phase lag at frequencies above 150 Hz were implemented as the stabilisation controllers. Experimental testing during the project revealed that the dominant disturbances expected to be incurred to the ISP if mounted on a moving vehicle would be at frequencies less than 5 Hz with a peak amplitude of less than 0.05 rad/s. Therefore, the integrator component of the controller was required in order to ensure adequate low frequency disturbance rejection performance. These loops had bandwidths of approximately 38 Hz and resonance peaks of less than 3 dB.

## V. Model Verification

The simulated model was verified experimentally as the final phase of the project. The simulated tracking step responses were compared to the implemented system's responses by comparing the error signals of the yaw and pitch tracking controllers of the simulation to the experimentally measured error signals of each channel's controller in response to control step commands. This is shown for the yaw channel in Fig. 9 and 10, where Ytc is the yaw tracking command, and Yte is the error signal. On the application of the step command both the simulated and implemented controllers overshoot the command by approximately 8 % before settling to the command after approximately 1.5 s. The simulation shows marginally greater oscillation than the physical test, however, it is hypothesised that the additional damping is due to disturbance torques exerted on the ISP by cable flexure which was not modelled in the system simulation. Overall it was concluded that the tracking controllers simulated performance matched the measure performance well.

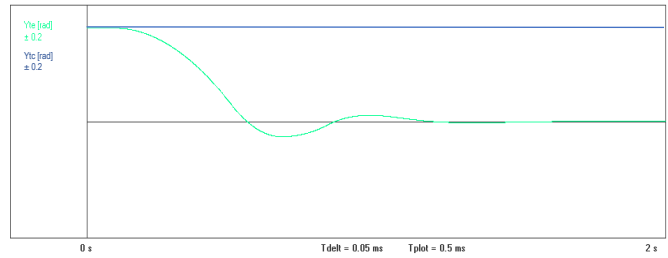

Fig. 9.   Simulated step tracking controller error on the yaw channel

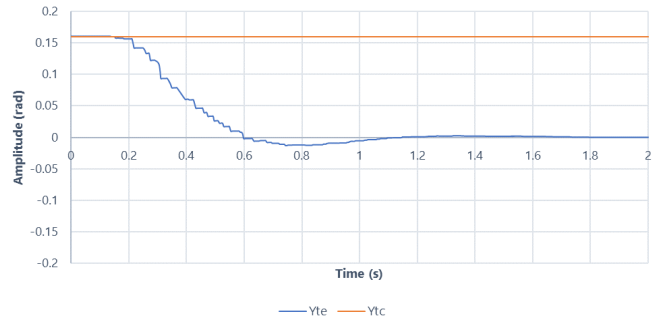

Fig. 10. Measured step tracking controller error on the yaw channel

Base motion isolation (BMI) of the ISP is defined below.

$$BMI = 20\log\left|\frac{Amplitude\ \omega_{np}}{Amplitude\ \omega_{nb}}\right| \quad (4)$$
$$where\ n \in \{x, y, z\}$$

BMI was evaluated in the worst-case gimbal orientation of $\theta = 45°, \psi = -45°$, by manually applying a base motion to the ISP, which was recorded using an MPU-9150 IMU attached to the system base and recording the LOS rates during the test. Several iterations of the test were repeated with the ISP in various orientations and for various base motion input profiles. Fig. 11 is indicative of the test results for the ISP. It shows the profile of a large base motion disturbance, $\omega_{yb}$, applied to the base and the associated LOS rates about the $y_p$ and $z_p$ axes. The median amplitude of the disturbance from $\omega_{yb}$ to $\omega_{yp}$ was reduced from approximately 28.93 °/s at 1.5 Hz to approximately 0.91 °/s, representing a BMI of -30.0 dB.

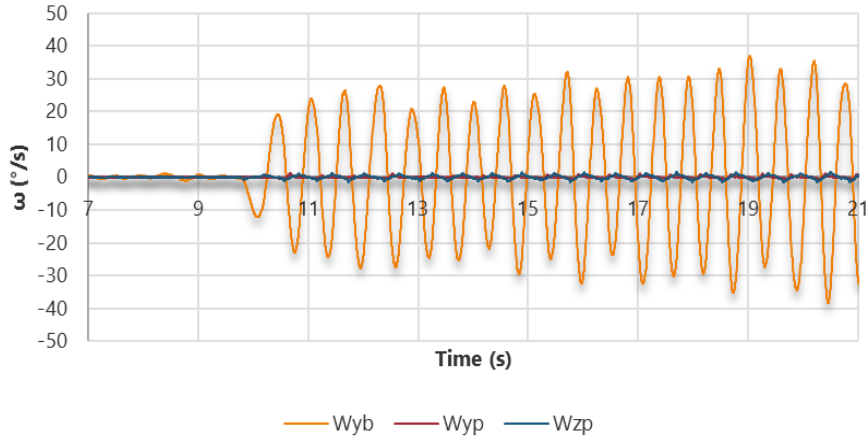

Fig. 11. Base motion attenuation of the ISP

For the same period, the median amplitude of $\omega_{zp}$ was approximately 1.39 °/s representing a cross-axis of approximately -26.3 dB. The validity of the simulation model for base motion tests was then established by repeating the experimental test conditions in the simulation environment. Fig. 12 shows a similar response to the hardware results under the test conditions used above: Inertial rate attenuation about the $y_p$ axis matches the test data within 8 %, whilst about the $z_p$ axis matches the test results within 3 %.

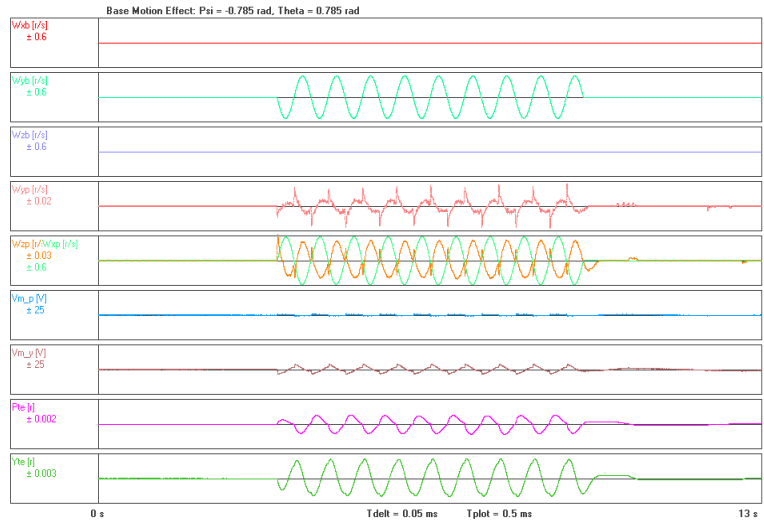

Fig. 12. Simulation responses to a 28.93 °/s sine signal about $y_b$ at 1.5 Hz

An approximate measure of angle jitter was evaluated for these tests by integrating the LOS rate signals about relevant axes. Overall, the performance achieved by the ISP across several tests is summarised by Table 1.

TABLE I. SUMARRY OF TESTED ISP PERFORMANCE PARAMERTERS

| Metric | Pitch Channel | Yaw Channel |
|---|---|---|
| BMI with disturbance axis aligned with attenuation axis | -34.4 dB (1.1 rad/s at ~2 Hz disturbance) | -31.5 dB (1.2 rad/s at ~2.5 Hz disturbance) |
| BMI with disturbance axis cross-coupled with attenuation axis | -30.0 dB (0.5 rad/s at ~1.5 Hz disturbance) | -26.3 dB (0.5 rad/s at ~1.5 Hz disturbance) |
| Maximum track overshoot | 11.6 mrad | 12.3 mrad |
| Maximum stationary (dynamic) jitter | 1.0 (2.35) mrad | 0.5 (2.52) mrad |

It can be seen from the test results above that BMI for the ISP for disturbances which are aligned with the attenuation axis are of the order of -32 dB which matches well with other platforms developed in contemporary literature [10], [11]. In addition, system jitter can also be seen to approximately be limited to the initial specification. The cross-coupled BMI is worse than the aligned BMI because the specific orientation depicted here is with initial yaw and pitch angles set at -45° and 45° degrees respectively. This is a scenario where cross-coupled BMI would be significant.

## VI. Conclusion

Overall, a two-axis stabilised platform was developed capable of stabilising a telescope modeller camera mount and achieving automatic tracking of the Moon. With reference to the main system specifications introduced earlier in this document, all major specifications were approximately met apart from the maximum allowable tracking error of 0.25 mrad. Tracking error achieved was 0.5 mrad and was due to the RPi computer not being able to process images of a high enough resolution to meet the specifications at a rate suitable for the tracking loops of the ISP control system, therefore requiring a limitation to be placed on the image resolution meaning that a 0.25 mrad error on the tracking system was not detectable to the system. In addition, the project ran marginally over budget with approximately R6700.00 being spent on development rather than the initial target of R5000.00.

However, automatic target tracking was achieved, and an expansion friendly mechanical design was implemented which can facilitate the inclusion of the Meade ETX-90 telescope easily. Classical controllers were designed and tested using a telescope modeller which inertially and geometrically modelled this telescope, and therefore it is not expected that the control algorithms will require much adjustment to effectively stabilise the ETX90 once it has been included. This system was experimentally shown to achieve similar performance to other ISPs developed for research purposes and to perform closely to the manner predicted by the simulation developed.